\documentclass[twocolumn,twoside,preprintnumbers,amsmath,amssymb,pacs]{revtex4}
\usepackage{epsfig}
\usepackage{graphicx}
\usepackage{dcolumn}
\usepackage{bm}

\usepackage{amssymb}

\usepackage{times}
\usepackage{epsf}

\usepackage{fancyhdr}

\usepackage{pslatex}

\begin{document}

\title{{\Large A STRATEGY TO STUDY CONFINEMENT IN QCD  }}

\author{Adriano DI GIACOMO}

\affiliation{PISA UNIVERSITY and INFN }

\received{on 24 March, 2006}

\begin{abstract}
The order and the universality class of the deconfining phase transition can provide insight into the mechanism of color confinement, in particular for $N_f=2$. The mechanism of confinement by monopole condensation is reviewed.


\end{abstract}
\maketitle
\thispagestyle{fancy}
\setcounter{page}{0}

\section{BASIC FACTS}
No free quarks are observed in Nature.  The upper limit to the ratio of the  abundance $n_q$ of quarks in ordinary matter to the abundance of nucleons $n_p$ 
is ${ n_q\over n_p}\le 10^{-27}$   \cite{PDG} to be compared to the expectation in the Standard Cosmological Model    ${ n_q\over n_p}\approx 10^{-12}$ \cite{okun} .

The cross section for inclusive production of quarks in p-p collisions, $\sigma_q$  is $\le 10^{-40}cm^2$ \cite{PDG} and should be of the order of the the total cross section i.e.$\approx 10^{-25}cm^2$ if quarks existed as free particles.  

An inhibition factor $\approx 10^{-15}$ implies that the only natural explanation of what is observed is that $n_q$ and $\sigma_q$ are exactly zero, protected by some symmetry.

A possible deconfining transition is then a change of symmetry, or an order-disorder transition, and can   not  be a cross-over.

 No experimental data exist on gluons . It is anyhow assumed that also gluons do not exist as free particles.  
 
 Confinement is defined as absence of colored particles in asymptotic states.
 
 If $QCD$ is the theory of strong interactions it should be able  to account for   confinement.
  
  In ref\cite{cabibbo}  the idea was first proposed that there could be a deconfining phase transition, at about the Hagedorn temperature $T_H$ from  the hadron phase to a plasma of quarks and gluons. The transition  is being searched in heavy ion collisions at CERN SPS, at Brookhaven RHIC and will be studied at higher energies at LHC.
  No smoking-gun signal for the formation of quark gluon plasma is known. One of the main difficulties 
  is to define operationally confined and deconfined. 
  
  A deconfining transition has been observed in numerical simulations of  $QCD$ on the lattice, in the absence of dynamical quarks (quenched theory).
  
  In general the thermodynamics of a  system is described by the partition function 
  \begin{equation}
  Z = Tr[ exp( -{H\over T})]
  \end{equation}
  with $H$ the hamiltonian .  For a field system it can be proved that the partition function is equal to the euclidean Feynman integral with the time axis compactified from $0$ to ${1\over T} $, and periodic boundary conditions for boson fields, antiperiodic for fermions.
  \begin{equation}
  Z = \int d\phi exp [-\int_0^{1\over T}d\tau \int d^3x L(\phi(\vec x,\tau))] 
  \end{equation}
 In lattice simulations this amounts to simulate the theory on a lattice $N_t N_s^3$  with the space extension $N_s \gg N_t$ ,the time extension. The temperature is then given by 
 \begin{equation} 
 T = {1\over {aN_t}}
 \end{equation}
 with $a$ the lattice spacing in physical units.
 Renormalization group at one loop gives the lattice spacing $a$ in terms of the physical scale $\Lambda_L$ of the lattice regularization scheme 
 \begin{equation}
 a = {1\over {\Lambda_L}} exp( {\beta\over {b_0}})
 \end{equation}
 with $\beta = {{2N}\over {g^2}}$  the natural variable of lattice and $b_0 = - {1\over {(4\pi)^2}}[{11\over 3}N- {2\over 3}N_f]$ the celebrated coefficient of the lowest order term of the $\beta$ function, which is negative (asymptotic freedom).
 
 From eqs(3) and (4) we get
 \begin{equation}
 T= {\Lambda\over{N_t}}exp( -{\beta\over {b_0}}) =   {\Lambda\over{N_t}}exp( {{2N}\over {|b_0|g^2}})
 \end{equation}
 As a consequence of asymptotic freedom  large $T$ corresponds to small $g$ ,or to order,
 small $T$ to large values of $g$ or to disorder. In usual thermodynamic systems there is disorder at high $T$ , order at small $T$.  This unusual behavior naturally leads to the idea of duality \cite{KW}\cite{KC}.
 
 Duality is a deep concept in statistical mechanics, string theory, field theory. It applies to systems in (d+1) dimensions with topologically non trivial spatial (d dimensional) configurations . The (1+1) dimensional Ising model is a prototype\cite{KC}.
 These systems have two equivalent complementary descriptions:
 
 1) A direct description in terms of local fields $\Phi(x)$ , in which  $\langle \Phi\rangle$ are the order parameters . We shall generically denote by $\mu$ the topologically non trivial spatial non local excitations.
 This description is convenient in the weak coupling regime $g\ll 1$.
 
 2) A dual description in which $\mu$ are local fields,  $\langle \mu \rangle$ the (dis)order parameters 
  and the fields  $\Phi(x)$ appear as non local excitations.  The typical coupling of the dual $g_D$ is
  $\approx {1\over g} $ and this description is convenient when $g_D \ll 1$ or $g\gg1$ , the strong coupling regime of the direct theory.
  
  The Ising model in (1+1) dimensions    is defined on a square lattice in terms of a local field $\Phi(x) $ which assumes the values $ \pm 1$ , with a ferromagnetic interaction between the nearest neighbors,
$H = -{\Sigma}_{\vec n,\vec n'} { \Phi}(\vec n).{\Phi}(\vec n')$  and $ Z[\beta ,\Phi] = exp(-\beta H) $.
  The model is exactly solvable. It has a phase transition at some known $\beta_c $ :  
  for $\beta\ < \beta_c $   the system is magnetized , i.e.   $\langle \Phi \rangle  \neq 0 $  ; for $\beta \geq \beta_c$  $\langle \Phi \rangle  =  0 $ and the system is disordered. If one of the two dimensions $n_0$ of the system is interpreted as time  and the other one $n_1$as space, the configurations with non trivial spatial topology are the kinks (antikinks) , which have spin $\Phi = -1$   for $n_1<  x_1$ ( $n_1 \ge x_1$) and $+1$   elsewhere at  fixed $n_0$.
  It can be shown that the operator $\mu $ which creates a kink also has values $\pm 1$ \cite{KC}\cite{CDL}, and that\cite{KC} 
  \begin{equation} 
  Z[\beta, \Phi]  = Z[ \beta^*, \mu]
  \end{equation}
   with 
 \begin{equation}
  sinh(2 \beta)  =   {1\over sinh(2 \beta^*)}
 \end{equation}
  The disordered phase in the direct description is ordered in the dual description. Duality maps the strong coupling disordered regime of the direct description into the ordered weak coupling regime 
  of the dual description and viceversa.
  This suggests to look for the dual description of $QCD$ in which the low temperature phase
  should be ordered , and to identify the topologically non trivial spatial configurations $\mu$
  and the symmetry of the dual description, i.e.  the disorder parameters $\langle \mu \rangle$.
  
  There are a number of candidate  dual excitations , namely monopoles \cite{tH81} , vortices\cite{tH78},
  AdS5 strings \cite{MA}.  
  
  It is fair to say that the dual excitations are not yet known , but something can be said about the dual symmetry which should be responsible for confinement.

\section{THE STRATEGY}
The facts described above suggest a strategy  for  investigating  confinement in $QCD$ :

I) Check on the lattice if the deconfining transition is indeed order-disorder, i.e. check if there is a natural explanation in terms of symmetry of the huge inhibition to  the existence  of free quarks.  Notice that the order of the deconfining phase transition can be studied without knowing the order parameter ,  by looking at scaling behavior of the specific heat , as we shall see below. 

II) Identify the symmetry responsible for confinement , and possibly the dual excitations.

Of course one has first to detect a deconfining transition on the lattice. This task proves to be equally complicated as in nature.

In the so called quenched approximation , or in pure gauge theory without dynamical quarks one can use the Polyakov line , which is the parallel transport in the time direction across the lattice. 
\begin{equation}
L(\vec x) =  Tr P exp(i\int_0^{1\over T}A^0(\vec x,t)dt)
\end{equation}

Due to periodic boundary conditions at finite $T$  the path is closed and due to the trace  $L$ is gauge invariant. 

Consider now the large distance behavior of the space correlator of  $L$  
\begin{equation}
D(\vec x) \equiv  \langle L^{\dagger}(\vec x) L(0) \rangle
\end{equation} 
On the one hand $ D(\vec x)$ is related to the static potential between a quark and an antiquark as
\begin{equation} 
V(\vec x) =  - T lnD(\vec x)
\end{equation}
On the other hand, by cluster property it behaves at large distances as
\begin{equation}
D(\vec x) \approx _{x\to \infty} |\langle L(\vec 0) \rangle|^2 + C exp(-{{\sigma x}\over T})
\end{equation}
It follows that if  $\langle L \rangle =0$
\begin{equation}
V(\vec x)\approx_{x\to \infty} \sigma x
\end{equation}
which means confinement of quarks . If instead $\langle L \rangle \neq 0$ then
\begin{equation}
V(\vec x)\approx_{x\to \infty} constant
\end{equation}
which means no confinement.
On the lattice a transition is observed at $T_c\approx 270 Mev $ from a phase in which $\langle L \rangle =0$ to a phase in which $\langle L \rangle \neq 0$.  The transition is first order for $SU(3)$ gauge theory,
$\langle L \rangle$ is the order parameter, $Z_3$ , the center of the group, is the symmetry.

In the presence of dynamical quarks $Z_3$ is not a symmetry, and therefore it can not be the symmetry responsible for confinement in nature. Nor can the string tension be an order parameter: even in the presence of confinement there is string breaking. When pulling apart a static $q-\bar q$ pair at some distance the system prefers to convert the potential energy into quark pair creation.

The situation for two flavor $QCD$ is shown in Fig[1] , where the phase diagram  is displayed. The two quarks are assumed for simplicity to have the same mass m \cite{DDP}. 
\begin{figure}[htbp]
\begin{center}
\includegraphics[width=8cm]{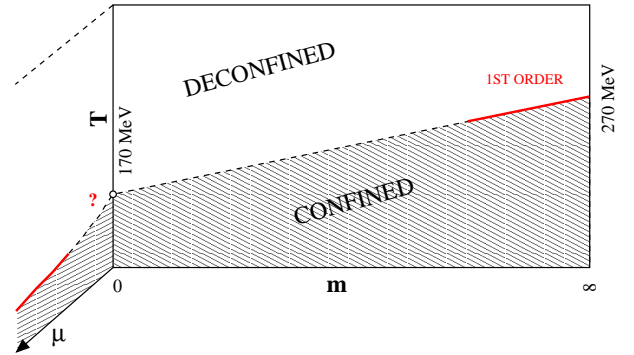}
\caption{The phase diagram of $N_f=2  QCD$. The transition line is defined by the maxima of the specific heat and of the susceptibility of the chiral order parameter. $m$ is the quark mass, $\mu$ the baryon chemical potential. }
\end{center}
\end{figure}

 At $m\approx 0$ chiral symmetry, which is broken at low temperature, is restored at $T_c\approx 170 Mev$ .  Above$T_c$  the chiral order parameter $\langle \bar \psi \psi\rangle$ vanishes. However at $m\neq 0$ chiral symmetry is not a symmetry any more. The transition line in Fig[1]  is defined by the maxima of 
 various susceptibilities, the specific heat, the susceptibility of the chiral order parameter, the susceptibility of the Polyakov line, which all coincide within errors. This coincidence means that 
 when the line is crossed a rapid change occurs in the energy  content, in the chiral order parameter,
 in the Polyakov line. Conventionally the region above the line is called deconfined phase, the region below it confined, but there is no way to test this statement operationally. This is analogous to what happens for heavy ion experiments. The problem would be solved if there were an order parameter
 for confinement.
 
 An analysis can be made of the chiral transition ($m=0$ ) \cite{PW}, assuming that the relevant 
 degrees of freedom at the transition are scalars and pseudoscalars. The order parameters are then the elements of the matrix:
 \begin{equation}
 \bar \Phi \equiv \Phi_{ij} = \langle \bar\psi_i(1+\gamma_5)\psi_j\rangle
 \end{equation}
  By one loop renormalization group analysis of the most general lagrangean compatible with chiral 
  symmetry, neglecting irrelevant terms , one looks for infrared stable fixed points, which are a necessary condition for a second order phase transition. If no such fixed points exist the transition must be first order. If one of such points exists, the transition can be second order.
  For $N_f \le  3$ no fixed point exists, so that the transition is first order, and this fits with lattice data\cite{B}.
  
  For $N_f = 2$ instead there are two possibilities:
  1) The transition is first order if the $\eta' $ mass $m_{\eta'} \to 0$ as  ${T\to T_c}$. Then  the transition is 
  first order also at $m, \mu \ne 0$ and there is no tricritical point at $\mu\ne 0$. $\mu$ here is the chemical potential. (See Fig[1])
   This possibility is consistent with the order disorder nature of the transition.
   
   2) The transition is second order. Then $m_{\eta'} \ne 0$ at $T_c$ , there is no transition at  $m\ne 0$
   but only a crossover , and a tricritical point at $\mu\ne 0$.  The crossover is incompatible with 
   an order disorder type of deconfining transition. It requires an unatural fine tuning in the theory as discussed in Section 1.
   
   It is therefore of fundamental importance to determine the order of the transition by lattice simulations.
   
   The tool to do that is a renormalization group based technique known as finite size scaling analysis.
   It consists in analyzing the volume dependence of susceptibilities as the spatial size of the system $L_s$ is sent large, which is governed by the critical indexes. Explicitly  for the specific heat  $C_V$ and for the susceptibility  $\chi$ of the order parameter one has \cite{KAR}\cite{DDP}
  \begin{equation}
   C_V - C_0 = L_s^{\alpha\over \nu} \Phi_C (\tau L_s^{1\over \nu}, mL_s^{y_h})
   \end{equation}
   and
   \begin{equation}
  \chi_{\langle \bar \psi\psi\rangle} - \chi_0 = L_s^{\gamma\over \nu} \Phi_ {\langle \bar \psi \psi \rangle}(\tau L_s^{1\over \nu}, mL_s^{y_h})
  \end{equation}
  Here  $\tau \equiv (1- {T\over T_c})$  and $\alpha$, $\gamma$ , $\nu$ and  $y_h$ are the relevant critical indexes in the standard notation. Their values for weak first order, second order $O(4)$ and $O(2)$ are listed in table I. $O(2)$ is included since the staggered formulation for the fermions which is used in the numerical simulations could break $O(4)$ to $O(2)$ at finite lattice spacing.
  \begin{table}[bt!]
\begin{tabular}{|c|c|c|c|c|c|}
\hline & $y_h$ & $\nu$ & $\alpha$ & $\gamma$ &
$\delta$\\
\hline $O(4)$ &  2.487(3) & 0.748(14) & -0.24(6) & 1.479(94) &
4.852(24)\\
\hline $O(2)$ & 2.485(3) & 0.668(9) & -0.005(7) & 1.317(38) &
4.826(12)\\
\hline $MF$ & $9/4$ & $2/3$ & 0 & 1 & 3\\
\hline $1^{st} Order$ & 3 & $1/3$ & 1 & 1 & $\infty$\\
\hline
\end{tabular}
\caption{Critical exponents.}\label{CRITEXP}
\end{table}
 
 The volume dependence  of the susceptibilities measured on the lattice simulations is then compared to Eqs. (15) and (16) to determine the critical indexes, or the order and the universality class of the transition.
 
 There are two scaling variables in the problem, $\tau L_s^{1\over \nu}$  and  $mL_s^{y_h}$  which makes the analysis difficult. Our strategy has been to keep one of the variables fixed  in turn and to observe the dependence on the other one\cite{DDP}.  
 
 First we have varied $m$ and $L_s$ by keeping  $mL_s^{y_h}$ fixed . To do that one has to assume 
 a value for $y_h$ , which we have first put equal to that of $O(4)$ and $O(2)$ which happen to be equal [See table I]  \cite{DDP} and then equal to that of first order. The expected scaling is then
 \begin{equation}
( C_V - C_0) /L_s^{\alpha \over \nu}  =  F_C (\tau L_s^{1 \over \nu})
\end{equation}
  and
\begin{equation}
(\chi_{\langle \bar \psi\psi\rangle} - \chi_0) / L_s^{\gamma\over \nu}  = F_ {\langle \bar \psi \psi \rangle}(\tau L_s^{1\over \nu})
\end{equation}
The  quantities on the left hand side of  Eqs(17) , (18)  should only depend on $\tau L_s^{1\over \nu}$ .
 The scaling for $O(4)$ is shown in fig[2] . If there were scaling all the curves in the same figure should coincide within errors.  The transition is not compatible with $O(4)$.
 Similar results are obtained for $O(2)$.  Preliminary data for the case in which  $mL_s^{y_h}$ is kept fixed assuming for $y_h$ the value for first order \cite{CDDP} seem instead compatible with the corresponding scaling  equations Eqs(17),(18)

 \begin{figure*}[htbp]
\begin{center}
\includegraphics[width=16cm]{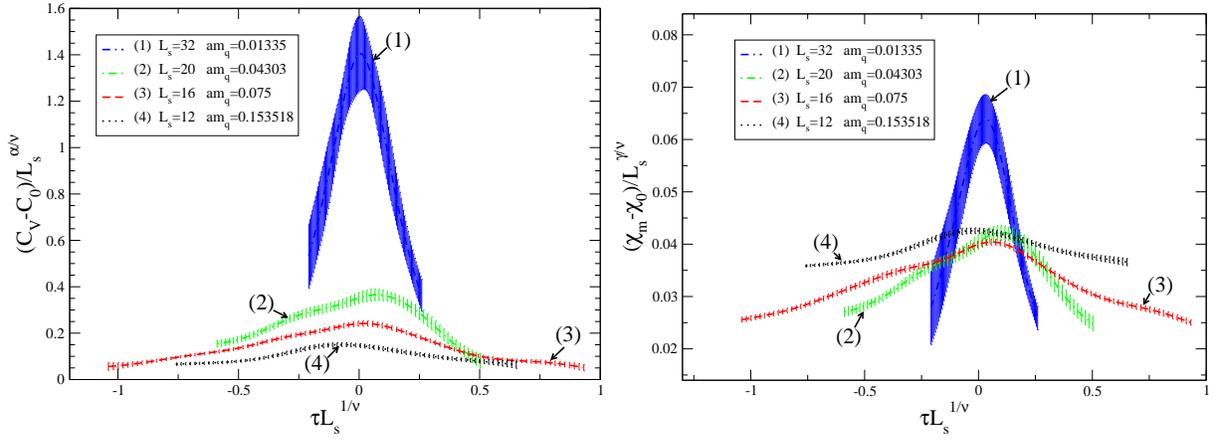}
\caption{Check of Eqs.(17), (18) assuming $O(4)$ second order. No scaling observed.}
\end{center}
\end{figure*}


 Another possibility is to go large with $L_s$ keeping the variable $\tau L_s^{1\over \nu}$ fixed. Here large $L_s$ means that the extension of the lattice is much bigger than the inverse mass of the pion,
 say at least 10 times. The following scaling laws are then expected
 \begin{equation}
 ( C_V - C_0) = m^{\alpha \over {y_h\nu}} G_C(\tau L_s^{1\over \nu})
 \end{equation}
 and
 \begin{equation}
 \chi_{\langle \bar \psi\psi\rangle} - \chi_0 = m^{\gamma  \over {y_h\nu}} G_{\langle \bar \psi\psi\rangle}(\tau L_s^{1\over \nu})
 \end{equation}
 These scaling laws can be checked both with  $O(4)$[$O(2)$] and with first order. The result is shown in Fig[3] and again favors first order.
  Simulations in which approximations of the algorithm are eliminated and data are obtained with an exact algorithm \cite{CDDP} confirm these results.
  
  The problem is numerically very demanding, with CPU times of the order of Teraflops . year.
  We plan however to repeat the analysis on larger lattices and with improved actions ,
  since the issue  "order-disorder" or crossover is fundamental.
  \begin{figure*}[htbp]
\begin{center}
\includegraphics[width=16cm]{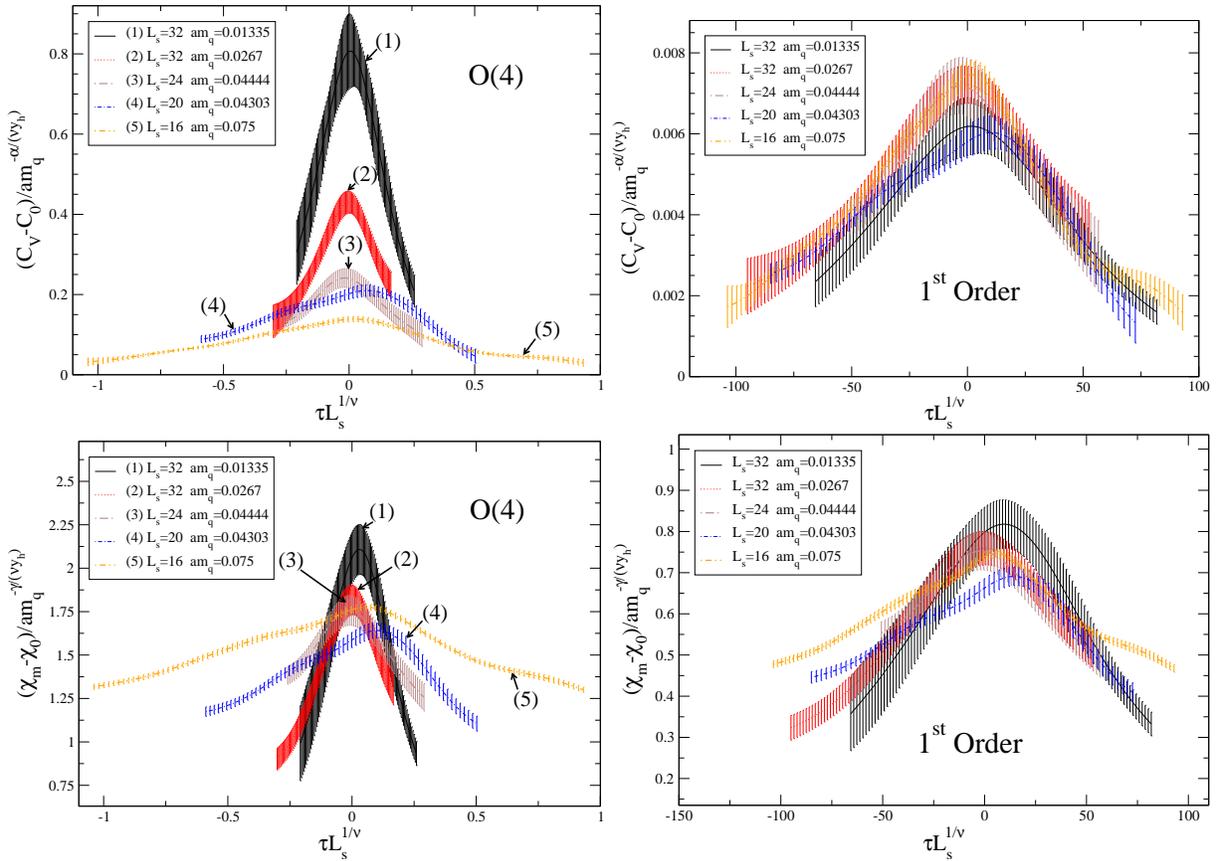}
\caption{Scaling Eqs. (19),(20) for second order $O(4)$ (left) and first order (right)}
\end{center}
\end{figure*}

 \section{THE DUAL EXCITATIONS. THE DUAL SYMMETRY}
 A popular candidate for dual excitations are vortices\cite{tH78}.  The number of vortices is a conserved quantity in (2+1) space time dimensions, not in (3+1) dimensions. Moreover in the presence of quarks  the symmetry $Z_3$ is explicitly broken. Gauge models with the same center have very different behavior at the deconfining transitions, and centerless systems like the gauge theory with gauge group $G_2$ exhibit confinement\cite{BERN}.
 We shall then concentrate on magnetically charged dual excitations (monopoles).
 A gauge invariant field-strength tensor $F^a_{\mu\nu}$ can be defined , the so called 'tHooft tensor\cite{tH74}, which is nothing but the field strength of the residual gauge symmetry after abelian projection. The index $a, (a=1,..,N-1)$ refers to the (N-1) different species of monopoles which 
 exist in $SU(N)$ gauge theory\cite{tH81} \cite{D}. The dual tensor is defined as usual 
 \begin{equation}
  \tilde F^a_{\mu,\nu} = {1\over 2}\epsilon_{\mu \nu \rho \sigma} F^a_{\rho \sigma}
  \end{equation}
  and  the magnetic current as
  \begin{equation}
  j^a_{\nu}  = \partial_{ \mu} \tilde F^a_{\mu \nu} 
  \end{equation}
   The magnetic current is identically zero if Bianchi identities are satisfied . It can be different from zero in a compact formulation of the theory , in terms of parallel transports like the lattice formulation. Monopoles do indeed exist on the lattice.
   The magnetic current is in any case conserved due to the antisymmetry of the 'tHooft tensor .
   \begin{equation}
   \partial_{\mu} j^a_{\mu} = 0
   \end{equation}
    A topological symmetry not related by Noether's theorem to the form of the Lagrangean.
    This symmetry can either be exact (Wigner) , and then the magnetic charge is defined and superselected, or be Higgs broken. In that case the ground state has no definite magnetic charge 
    and the system is  a dual superconductor : magnetic charges condense like Cooper pairs (electric charges) do in an ordinary superconductor.
     This symmetry is a good candidate symmetry for confinement . In the confined phase ($T\leq T_c$) the vacuum behaves as a dual superconductor. The Chromoeletric field acting between a $q-\bar q$ pair is channeled by dual Meissner effect into Abrikosov flux tubes . Since their energy is proportional to the length of the tube, i.e. to the distance between the pair , this explains confinement. Above $T_c$ the
     Higgs phenomenon disappears , magnetic charge is superselected, and confinement disappears.
     Such a mechanism for confinement was proposed long ago in refs\cite{MAN}\cite{tH1}.
     The idea has been  directly checked on the lattice. 
     An operator  $\mu ^a(x)$ can be defined\cite{DP}\cite{FM} \cite{DLMP}, which creates a monopole of species $a , (a=1,..,N-1)$ at  $x$. In the deconfined phase one expects $\langle \mu^a \rangle =0$
     if there is superselection of magnetic charge.   $\langle \mu^a \rangle \neq 0$ in the deconfined phase  indicates Higgs breaking and dual superconductivity of the vacuum.
     
     The operators  $\mu ^A(x)$ are the lattice transcription of the continuum operators
     \begin{equation}
  \mu^a (\vec x,t) =  exp[ i\int d^4 y\vec E^a_{\perp}(\vec y,t) b_{\perp}(\vec y - \vec x)]
    \end{equation}
     where $b_{\perp}(\vec y - \vec x)$  is the vector potential produced in $\vec y$ by a monopole sitting at
     $\vec x$, $\vec E^a_{\perp}$ is the color component of the chromoelectric field along the direction of the residual gauge symmetry which identifies the monopole \cite{DLMP}\cite{D}. Since  $\vec E^a_{\perp}$ is the conjugate momentum to $\vec A^a_{\perp}$  the operator $\mu^a (\vec x,t)$ simply adds to $\vec A^a_{\perp}$ the field of the monopole.It is easy to show that \cite{DLMP}
     \begin{equation}
      \langle \mu^a \rangle = {{\int dA exp[-\beta (S+\Delta S)]}\over {\int dA exp[-\beta S]}}
      \end{equation}
      It is then convenient to use, instead of $\langle \mu^a\rangle$  the quantity
      \begin{equation}
      \rho^a \equiv {{\partial ln(\mu^a)}\over {\partial \beta}} = \langle S \rangle_S - \langle (S + \Delta S)\rangle_{(S+\Delta S)}
      \end{equation}
      Since Eq(25) implies that $\langle \mu^a \rangle = 1$ at $\beta=0$ it follows by integration of Eq(26)
      that
      \begin{equation}
      \langle \mu^a \rangle  = exp[ \int^{\beta}_0 \rho(\beta') d\beta']
      \end{equation}
      Lattice simulations show the following:
      
     1)  For   $T < T_c$  or  $\beta<  \beta_c$  , $\rho$ tends to a finite limit as the spatial size 
     of the lattice       $L_s \to \infty$. In practice 
     it becomes volume independent when the lattice size is larger than $1fm$. This implies by use of
     Eq(27) that    $\langle \mu^a \rangle \neq 0$  below the deconfining transition and hence that there
      is   dual superconductivity
      
     2) For  $T > T_c$ or $\beta > \beta_c$ , i.e. above the deconfining transition as $\L_s \to \infty$
     \begin{equation}
     \rho^a \approx  - c L_s + c'
     \end{equation}
     with $c > 0$ implying by use of Eq(27) that  $\langle \mu^a \rangle = 0$ in the thermodynamic limit,
     or superselection of the magnetic charge.
     3)  For $T\approx T_c$ or $\beta \approx \beta_c$ the scaling law holds
     \begin{equation} 
     \langle \mu^a \rangle \approx L_s^{\kappa}\Phi_{\mu} (\tau L_s^{1\over \nu},m L_s^{y_h})
     \end{equation}
     If $L_s m_{\pi} \gg 1$ the above formula becomes
     \begin{equation}
     \langle \mu^a \rangle \approx m^{\kappa\over {y_h}}f_{\mu} (\tau L_s^{1\over \nu})
     \end{equation}
     which implies the scaling law
     \begin{equation}
     {\rho^a \over {L_s^{1\over \nu}}} \approx  F(\tau L_s^{1\over \nu})
     \end{equation}
     The scaling law Eq(31) can be used to determine the critical index $\nu$, and has the remarkable property of being $m$ independent .  It has been successfully checked for $U(1)$ , $SU(2)$, $SU(3)$ pure gauge theories\cite{DP}\cite{DLMP} \cite{DLMP1} giving the correct critical index . For $N_f=2 QCD$ it gives a very good scaling with the critical index corresponding to first order\cite{DDLPP}. This is consistent with the results for the susceptibilities described in Sect 2.
     \section{CONCLUSIONS}
     We have argued that the only natural explanation of experimental data on confinement is that
     confinement  is related to  a symmetry, and therefore that the deconfining phase transition is an order disorder transition, and not a crossover.
     
     The primary goal is then to check this statement, which amounts to determine the order of the deconfining transition.For that the study of specific heat is specially convenient. Preliminary results on $N_f=2$ $QCD$ around the chiral transition are  consistent with it.
     
     Only the existence of an order parameter allows to define operationally confined and deconfined .
     
     No definite statement can be made about the dual excitations of QCD. 
     
     Something can be said on their symmetries : they carry magnetic charge. Vacuum is a dual superconductor in the confined phase,
     and makes a transition to normal at deconfinement .
     
      At higher temperatures magnetic charge is superselected.
      
     We are working to refine and to assess the results reported by simulations on larger lattices and
     with improved action.
     
      This work was partially supported by M.I.U.R. Progetto:Teoria e fenomenologia delle interazioni fondamentali





\end{document}